\documentstyle[12pt,aasms4]{article}
\def \Lya{{Ly-$\alpha$}}

\def \Lyf{\Lya\ forest}
\def \Lyc{\Lya\ continuum}
\def \Ha{H$\alpha$}

\def \cf{cf.}

\def \etal{{\it et~al.}}
\def \ie{{\it i.e.,}}

\def \ksM{km~s$^{-1}$~Mpc$^{-1}$}
\def \apj{ApJ}

\def \cgs{ergs cm$^{-2}$ s$^{-1}$}
\def \cgsa{ergs cm$^{-2}$ s$^{-1}$ \AA$^{-1}$}
\def\deg{\ifmmode {^{\circ}}\else {$^\circ$}\fi}
\def\hr{$^{h}$}
\def\min{$^{m}$}
\def\secper{\ifmmode \rlap.{^{s}}\else $\rlap{.}{^{s}} $\fi}
 
\lefthead{Weymann et al.}
\righthead{Keck Spectroscopy and NICMOS Photometry of a Redshift $z = 5.60$ Galaxy}

\begin{document}

\title{Keck Spectroscopy and NICMOS Photometry of a Redshift $z = 5.60$ Galaxy\footnotemark[1]}

\author{Ray J.\ Weymann\altaffilmark{2,3}} 
\affil{E-mail:~rjw@ociw.edu}

\author{Daniel~Stern\altaffilmark{4}, Andrew~Bunker\altaffilmark{4},
Hyron~Spinrad\altaffilmark{4}}
\affil{E-mail:~dstern,abunker,hspinrad@bigz.berkeley.edu}

\author{Frederic H.\ Chaffee\altaffilmark{5}}
\affil{E-mail:~fchaffee@keck.hawaii.edu}

\author{Rodger I.\ Thompson\altaffilmark{6}}
\affil{E-mail:~thompson@as.arizona.edu}

\and
\author{Lisa J. Storrie-Lombardi\altaffilmark{3}}
\affil{E-mail:~lisa@ociw.edu}

\author{[ to appear in the Astrophysical Journal Letters ]}

\footnotetext[1]{ Optical data
presented herein were obtained at the W.M.~Keck Observatory, which is
operated as a scientific partnership among the California Institute of
Technology, the University of California and the National Aeronautics and
Space Administration. The Observatory was made possible by the generous
financial support of the W.M.~Keck Foundation. 
The near-infrared observations were obtained with the 
Near-Infrared Camera and
Multi-Object Spectrometer on the 
NASA/ESA Hubble Space Telescope
which is operated by AURA Inc., under contract with NASA.}
\altaffiltext{2}{Visiting Prof., Lick Observatory, Santa Cruz, CA 95064}
\altaffiltext{3}{Observatories of the Carnegie Institution of Washington, 
813 Santa Barbara Street, Pasadena, CA 91101}
\altaffiltext{4}{Astronomy Dept., 601 Campbell Hall, U. C. Berkeley, Berkeley, CA 94720}
\altaffiltext{5}{W.~M.~Keck Observatory, 65-1120 Mamalahoa Hwy., Kamuela, HI 96743}
\altaffiltext{6}{Steward Observatory, The University of Arizona,
Tucson, AZ 85721}

\begin{abstract}		               
We present Keck LRIS spectroscopy along with
NICMOS F110W ($\sim$ J) and F160W ($\sim$ H)
images of the galaxy HDF~4--473.0 (hereafter 4-473) in the 
Hubble Deep Field,
with a  detection of an emission line consistent with \Lya\
at a redshift of $z=5.60$. Attention to this object as a high redshift
galaxy was first drawn
by Lanzetta, Yahil and Fernandez-Soto and appeared in their initial
list of galaxies with redshifts estimated from the WFPC2 HDF photometry.
It was selected by us for spectroscopic observation, along with others
in the Hubble Deep Field, on the basis of the
NICMOS F110W and F160W and WFPC2 photometry. For H$_0$ = 65 {\ksM} 
and q$_0$ = 0.125, 
use of simple evolutionary models along with the 
F814W ($\sim$ I), F110W, and F160W magnitudes allow us to 
estimate the star formation rate ($\sim 13 M_\odot/$yr). 
The colors suggest a reddening of
E(B-V) $\sim$ 0.06. The measured flux in the \Lya\ line is approximately
$1.0~ \times 10^{-17}$ \cgs\ and the restframe equivalent width, 
correcting for the
absorption caused by intervening H~I, is approximately 90 \AA. 
The galaxy is compact and regular, but resolved, with an observed 
FWHM of $\sim0.44''$.
Simple evolutionary models can accurately reproduce the colors
and these models predict the \Lya\ flux to within a factor of 2.  Using this
object as a template shifted to higher redshifts, we calculate
the magnitudes through the F814W and two NICMOS passbands for galaxies at
redshifts $ 6 < z < 10$.
\end{abstract}

\section{Introduction}

There have recently been several programs aimed at discovering galaxies
at very high redshifts, utilizing the strong effects on broadband 
colors arising from the \Lyf\  and
Lyman continuum absorption,  
(Lanzetta \etal\ 1996; Madau \etal\ 1996; Steidel \etal\ 1996;  Spinrad \etal\ 1998;
Fernandez-Soto \etal\ 1998),
imaging to detect \Lya\ emission through narrow
band filters (Hu \etal\ 1998), Fabry Perot imaging 
(Thommes \etal\ 1998),
long  slit ``serendipitous'' \Lya\ searches 
(Dey \etal\ 1998; Hu \etal\ 1988),
and slitless spectroscopic searches (Lanzetta 1998). 
The ability of the HST NICMOS (Thompson \etal\ 1998a)  to obtain very deep images
over the range $0.8\mu$ to $1.8\mu$ offers an additional tool for discovering
faint high redshift candidates.  We report here the detection of the
first such galaxy incorporating this NICMOS data, and describe some of its properties.

\section{Target Selection, NICMOS Images and Photometry}

During the first NICMOS Camera 3 HST Campaign in January 1998, very
deep images were taken of a portion of the WFPC2 HDF Chip 4 in both the
F110W and F160W filters, corresponding approximately to the J and H
bands. A catalog of the galaxies found on these images and discussion of
the contents of this catalog is presented elsewhere 
(Thompson \etal\ 1998b). 
Preliminary inspection of these images revealed that the galaxy 4-473
in the Williams \etal\ (1996) catalog 
was relatively bright in F110W and F160W,
distinctly fainter in the F814W (I) band and not visible in F606W (V).
Although we were unaware of it at the time, this object had already been
noted by Lanzetta, Yahil \& Fern\'andez-Soto (1996) as a high redshift candidate, with a photometrically
estimated redshift from the F814W and F606W images of 6.8.
The negative F110W--F160W (AB) NICMOS color, together with the moderately
red (positive) F814W--F110W AB color made it a very strong candidate for spectroscopic
observations to confirm it as a high redshift galaxy.
Indeed, combined with its
non-detection in KPNO IRIM IR imaging (Dickinson \etal\ 1998) the revised
photometric redshift of 5.64 by 
Fern\'andez-Soto, Lanzetta \& Yahil 1998  is remarkably close to what we have 
observed spectroscopically.

To measure colors on a consistent basis, we have measured $1.0''$
diameter aperture magnitudes (AB scale used throughout) and obtain 
values of $27.12~\pm~0.19$ in F814W, $26.64~\pm~0.04$ in F110W and
$26.86~\pm~0.04$ in F160W. 
The S/N of the corresponding F606W image is $<$ 1, and 
the lower limit on the F606W - F814W color is 2.31 
from Williams \etal\ (1996).
A mosaic of  $3''\times3''$ regions
around 4-473 in these 4 passbands is shown in Figure 1. 

\section{Spectroscopic Observations of 4-473}
Although 4-473 is extremely faint, recent successes in detecting
\Lya\ emission from very high redshift galaxies (Dey \etal\ 1998;
Hu \etal\ 1998) encouraged us to attempt spectroscopic observations.
Accordingly we observed the HDF on 25 February 1998 UT using the
slitmask spectroscopic mode of the Low Resolution Imaging Spectrometer
(LRIS; Oke \etal\ 1995) at the Cassegrain focus of the
Keck II Telescope.  The $1.5''$ wide slitlet containing 4-473
was $\approx 24''$ long, allowing for effective sky removal. 
The observations were made with the 400 l/mm grating ($\lambda_{\rm blaze}
\approx 8500$ \AA; $\Delta \lambda \approx 11$ \AA).  
The telescope was offset $2.5''$ along the slit
between each 1800 second exposure to
facilitate removal of the fringing in the near--IR regions of the
spectrogram.  A total of 9000 seconds of integration was obtained, and the
reduced spectrogram revealed a faint emission line at $\approx 8030$
\AA\ at the position of 4-473 on the slitlet.  We therefore
reobserved 4-473 for an additional 5400 seconds on 27 June 1998 UT with
LRIS in longslit mode, utilizing the relatively bright star
HDF~4--454.0 at the center of HDF chip 4 as a control for both
spectrophotometry and spectroastrometry.  These observations were
obtained with a $1''$ slit and the 400 l/mm grating and 
confirmed the emission line detected in February.

Data reductions were performed using the {\it IRAF} package and
followed standard slit spectroscopy procedures.  Wavelength calibration
was performed using a NeAr lamp, employing telluric lines to adjust the
zero--point.  Flux calibration employed a sensitivity function derived
from 20,21 January 1998 UT observations of G191B2B and HZ~44 (Massey
\etal\ 1988; Massey \& Gronwall 1990) and was adjusted to ensure F814W
magnitudes derived from spectrophotometry of the brighter serendipitous
objects HDF~4--460.0 (February 1998 data) and HDF~4--454.0 (June 1998
data) matched the Williams \etal\ (1996) imaging photometry.  Our
composite spectrogram, presented in Fig.~2, reveals a robust detection
of an emission line at 8029 \AA\, with an integrated flux of $\approx
1.0 \times 10^{-17}$ \cgs.

\section{Identification of the Emission Line as \Lya}
 
As discussed below, we believe the only reasonable identification 
for the emission line
in 4-473 is \Lya. Other possible alternatives are
\Ha, [O~III] 5007, and
[O~II] 3727; [O~III] 4960 or H$\beta$ are unlikely since
[O~III] 5007 would also be
detected. (By chance, the slitlet for the February observations
passed through the galaxy HDF 4--460.0, which has a redshift of
z $\sim$ 0.68 and has three lines at similar observed 
wavelengths as the line in 4-473.)
If the line were \Ha, the restframe colors of 4-473 would be unlike 
any galaxy of which we are aware. 
Identification of the line as [O~III] 5007 itself might
be possible since the corresponding [O~III] 4960 and H$\beta$ lines would be weaker
and in a region of strong OH night sky emission, but the corresponding
[O~II]3727 would fall in a region uncontaminated by night sky emission
and is not present to a very low flux level. By contrast, the [O~II] 3727
line in the spectrum of HDF~4--460.0 is readily observed and is very strong.

The colors may also be used to rule out with  high probability an
identification as [O~II] 3727. To make this latter statement quantitative 
we have
used the 6 template spectra of star--forming galaxies assembled by
Calzetti \etal\ (1994). These
templates are each sets of star-forming galaxies observed with both IUE and
from the ground, and grouped into 6 sets characterized by different amounts
of internal reddening.\footnotemark[7]
Shifting these templates to the redshift implied if the line 
were identified
as [O~II] 3727, the F606W - F814W colors range from
0.39 to 0.79, \ie\ 
much {\it bluer} than the observed limit of $> 2.31$. 
Moreover, for the
reddest of the templates, the F814W--F110W and F110W--F160W colors are
much  {\it redder} than the observed colors: 0.79 and 0.64
compared to the observed values of +0.48 and -0.22, respectively. Galaxies
with old stellar populations, in which a hidden source of ionizing radiation
might produce the [O~II] 3727 emission encounter similar difficulties.
\footnotetext[7]{We thank Dr. Calzetti for
kindly making available these templates to us in digital form.}

An additional argument in favor of the \Lya\ identification comes from
the profile of the line itself. As seen in Figure 2, the line is
asymmetric, with absorption on the blue side, just as anticipated
from absorption by local H~I and/or a dense \Lyf\ (\cf\ Dey \etal\ 1998). 
In the following we assume therefore
that the \Lya\ identification is correct, and the wavelength at the 
peak of the line emission implies a redshift of 5.603 $\pm$ 0.002.
 
\section{Discussion}
 
HDF~4--473 is not the galaxy with the highest reported redshift. 
A  galaxy of slightly
higher redshift has been reported in a serendipitous long slit exposure
by Hu \etal\ (1998), and no doubt more high redshift galaxies will be
forthcoming from the various approaches described in the Introduction
(\cf\ Lanzetta 1998). 
However, the combination of the WFPC2 and NICMOS images allow us to
make some estimates of the star formation rate as well as the
reddening.

\subsection{Empirical Estimate of Continuum Level and Slope}
We have been unable to detect a continuum in our spectroscopic 
observations,
since only a fairly limited portion of the spectrum redward of \Lya\
is free of strong OH emission. 
To estimate the continuum level we use a
semi-empirical model in which we assume a power law of the form
$$ F(\lambda) = F_0 \times (\lambda/1.6\mu)^{\beta} $$
and impose the absorption due to the \Lyf\ and \Lyc\ calculated recently
by Madau, Pozzetti \& Dickinson (1998). We then determine the values
of $F_0$ and $\beta$  as well the flux in the \Lya\ emission line from
the F160W, F110W and F814W magnitudes. We regard the flux in the \Lya\ emission
line as a parameter to be determined
since our absolute flux measurement is somewhat uncertain and
the \Lya\ flux makes a significant contribution to the total counts in the
F814W filter. The magnitudes are very accurately reproduced for this redshift
with $\beta$ = -2.40, a \Lya\ flux of $1.2~\times~10^{-17}$ \cgs\ and a continuum flux
at the \Lya\ line of $4.0~\times~10^{-20}$ \cgsa. This implies a restframe equivalent
width of about 45~\AA\ for the unobscured portion of the line, and about
90~\AA\ if, as appears to be the case from the line profile, half of the line
is absorbed away.

\subsection{Evolutionary Models, Star Formation Rate and Reddening}
We  construct evolutionary models as follows: 1) We use
the latest version of the GISSEL model (Bruzual \& Charlot 1993; 
Leitherer \etal\ 1996) for a Salpeter IMF and
$125M_\odot$ upper limit cutoff. 2) The Madau \etal\ (1998) Lyman absorption model
is used.
 3) We use the Calzetti \etal\ (1994) 
reddening law in which we have set the ratio, R, of reddening to extinction 
to 3.1. 4) We assume a constant rate of star formation extending from the observed epoch
over some period $\delta T$ years. 
5) We fix the flux of \Lya\ at $1.2~\times~10^{-17}$ \cgs\ on the
basis of the semi-empirical determination above. 6) We
adopt the  cosmological parameters H$_0 = 65$
\ksM and q$_0 =0.125$.

We have explored models in which we vary the reddening (characterized by E(B-V)),
the rate of star formation, and the duration of the star formation epoch, $\delta T$.
For a given reddening and star 
formation duration, we adjust the star formation
rate to force agreement with the F160W magnitude. The resultant F110W--F160W
and F814W--F110W
colors are rather insensitive to the star formation duration, and even the inferred
star formation rate is not strongly dependent upon $\delta T$ over the range
$20<\delta T <80$~Myr. We adopt in the following 
$\delta T$ = $5~\times~10^7$ years.
The F110W--F160W and F814W--F110W colors can be exactly matched with E(B-V) = 0.06,
and a star formation rate of $13 M_\odot/$yr. 

Estimates on the limits for the
reddening and corresponding star formation rates are approximately E(B-V)
$\sim0.00$, and SFR $\sim 8 M_\odot/$yr and E(B-V) $\sim$ 0.12 and SFR $\sim 19
M_\odot/$yr. These reddening and star formation rate estimates are
somewhat lower than those recently obtained for galaxies at 
z = 5.34  and z = 4.92
by  Armus \etal\ (1998) and Soifer \etal\ (1998).
At a wavelength of $\lambda9900$\AA, corresponding to a rest wavelength
of $\lambda1500$\AA, the model predicts a flux of $2.6 \times 10^{-20}$ \cgsa. 
Correcting for the extinction,  our adopted cosmological model
implies a luminosity of $1.4 \times 10^{41}$ ergs sec$^{-1}$ \AA$^{-1}$, which
agrees well with the relation between star formation rate and UV luminosity
proposed by Madau, Pozzetti \& Dickinson (1998). For a cosmology with 
H$_0$ = 50 {\ksM} and q$_0$=0.5 the corresponding luminosity would be lower
by a factor of 1.8.
We do not, of course, claim that these model properties are well determined, given
all the uncertainties inherent in the assumptions listed above. 
However, additional support comes from the {\it predicted}
\Lya\ flux. If subjected to the same reddening as the continuum
and if 50\% is absorbed (as seems reasonable based upon the 
asymmetric profile), this yields a flux of $1.8 \times 10^{-17}$ \cgs, 
in rough agreement with that observed. 
The modest excess in the predicted flux over
our best estimate for the observed value can be ascribed to destruction
by dust of the multiply scattered \Lya\ and/or by variations in the IMF to
which the \Lya\ flux is moderately sensitive.

\subsection{Morphology}
Inspection of the HST images reveals that 4-473 has a regular,
compact morphology. We note that the other $z>5$ galaxies for which
imaging exists either have multiple components (Spinrad \etal\ 1998) or
are resolved even from the ground (Dey \etal\ 1998).

In F160W the galaxy has FWHM=0.44$''$. Comparison with a PSF star
shows that 4-473 is clearly resolved (FWHM$_{\rm PSF}$=0.2$''$)
with a deconvolved half-light radius of 0.2$''$ (1.4~kpc), comparable
to that found for many of the $z\approx 3$ Lyman-break galaxies
(Giavalisco \etal\ 1996). The major/minor axial ratio is measured to be
$\frac{a}{b}=1.15\pm0.04$ in F160W.  Neither an exponential disk nor a
de~Vaucouleurs $r^{1/4}$ law fit the radial profile well. A two-component
model suggests that the disk may dominate, as with other ``spheroidal''
objects in the HDF (Marleau \& Simard 1998). The characteristic disk
scale length is $r_{0}\approx 2.4$~kpc.

There is a nearby brighter object, HDF~4--497.0, which Bunker \etal\
(1998) have shown to be a foreground system associated with the nearby
Lyman-break galaxy, HDF~4--555.1 (``the Hot Dog'' at $z=2.80$,
Steidel \etal\ 1996). There appears to be a diffuse structure between
HDF~4--497.0 and 4-473, the nature of which is uncertain. As this feature
is still visible in the F606W band, it is likely not part of the
higher-redshift system.

\subsection{Extrapolation to Higher Redshifts} 
Using the ``best fit'' model
above for the star formation rate and reddening, we can then predict the
magnitudes a galaxy like 4-473 would have at higher redshifts. A more
complete discussion is given in Thompson \etal\ (1998b). 
For H$_0$ = 65 {\ksM} and
q$_0$ = 0.125 and keeping the inferred \Lya\ 
luminosity fixed, we obtain the values in Table~1. 
For H$_0$ = 50 {\ksM} and q$_0$ = 0.5, the decrease in brightness with 
increasing redshift is less than for the H$_0$ = 65 and q$_0$ = 0.125 
cosmology assumed in Table 1. Relative to the magnitudes
in Table 1 the galaxy would be 0.04 magnitudes brighter at z = 6 
and 0.33 magnitudes brighter at z = 10.
Evidently, near IR imaging with NICMOS is, and will continue to be, 
a powerful tool for the study of high redshift galaxies out to z $\sim$ 10.

\acknowledgments{The near-infrared observations are supported 
by NASA grant NAG5-3042 to the NICMOS instrument definition team. 
RJW thanks
the W.M.~Keck Observatory and the Lick Observatory for 
their hospitality during the
period when this work was carried out and 
P. McCarthy and D. Koo for useful
discussions. AB and LJSL gratefully acknowledge financial support 
from NICMOS postdoctoral positions.}

\newpage
 
\begin{figure}
\figurenum{1}
\plotone{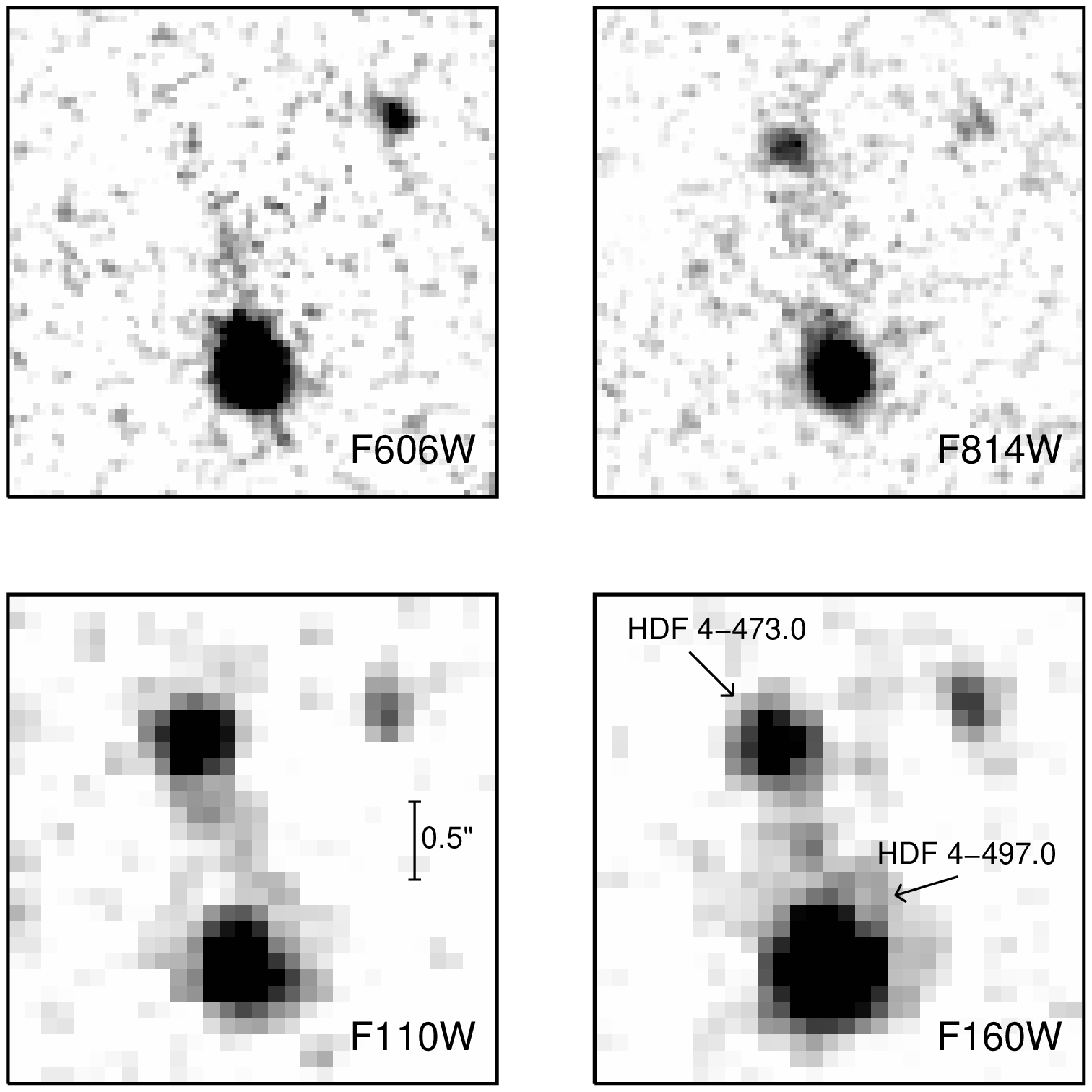}
\caption{ Mosaic
around HDF~4--473.0 in the F606W ($\sim$V), F814W ($\sim$I), 
F110W ($\sim$J), and F160W ($\sim$H) passbands.
AB $1''$ diameter aperture magnitudes are $>$29.4, 27.1, 26.6, and
26.9, respectively.
4-473 is located at $\alpha$ = 12\hr36\min45\secper902, $\delta$ =
62\deg11\arcmin58\farcs21 (J2000).  Images shown are $3''$ on a
side. Bunker \etal\ (1998) have shown HDF~4--497.0 to be at $z=2.80$.}
\label{hdfimage}
\end{figure}

 
\begin{figure}
\figurenum{2}
\plotone{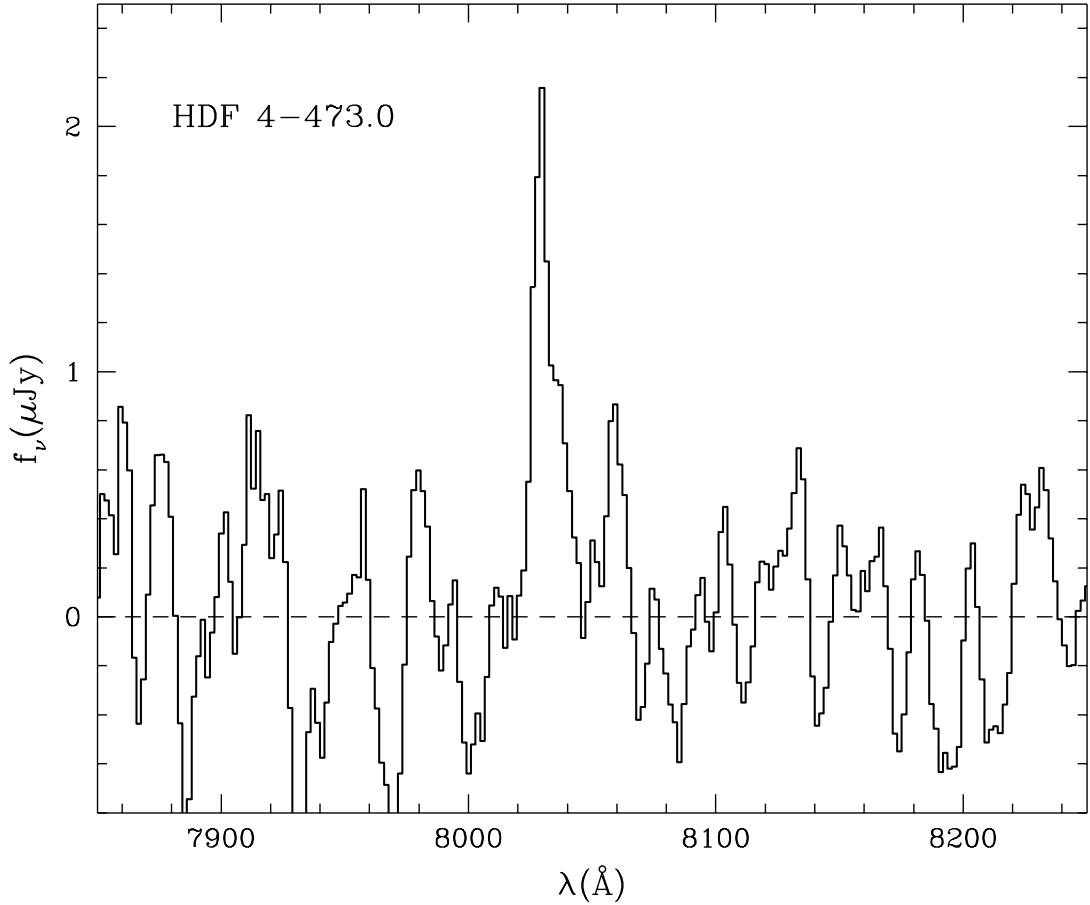}
\figcaption{
Spectrum of HDF~4--473.0 showing the emission line at 8029 \AA.  The
asymmetric line profile and broad band colors are consistent with this
being Ly$\alpha$ at $z=5.60$.  The total exposure time is 14,400 seconds,
and the spectrum was extracted with a $1.7''$ aperture.  The spectrum
has been smoothed using a 5-pixel boxcar filter. }
\label{hdfspec}
\end{figure}
 
\newpage
\begin{deluxetable}{ccccccccc}
\tablenum{1}
\tablewidth{0pc}
\tablecaption{Predicted AB magnitudes for Higher Redshift Galaxies}
\tablehead{
 \colhead{z} & \colhead{Flux} & \colhead{F160W} & \colhead{F110W} & \colhead{F814W} & \colhead{F606W} & \colhead{K} & \colhead{F110W-F160W}  &   \colhead{F814W-F110W} \\ 
 & \colhead{(Ly$\alpha$)} &  &  & & & & &    }
\startdata
 5.6 &  1.2e$-$17  &   26.86 & 26.64 & 27.13 & 30.09 & 27.07 &  $-0.22$  &  $+0.49$ \\
 6.0 &  9.6e$-$18  &   26.98 & 26.85 & 27.77 & 32.10 & 27.19 &  $-0.14$  &  $+0.92$ \\
 7.0 &  6.3e$-$18  &   27.21 & 27.33 & 31.36 & ---   & 27.39 &  $+0.12$  &  $+4.03$ \\
 8.0 &  4.3e$-$18  &   27.44 & 27.93 & ---   & ---   & 27.59 &  $+0.49$  &   --- \\
 9.0 &  3.1e$-$18  &   27.63 & 28.63 & ---   & ---   & 27.76 &  $+1.00$  &   --- \\
10.0 &  2.3e$-$18  &   27.81 & 29.66 & ---   & ---   & 27.93 &  $+1.85$  &   --- \\

\enddata
\label{t1}
\end{deluxetable}


\newpage
\begin{thebibliography}
\bigskip

\bibitem{}{Armus, L., Matthews, K., Neugebauer, G., \& Soifer, B.T. 1998,
ApJ, in press}

\bibitem{}{Bruzual, G.A. \& Charlot., S. 1993, ApJ, 405, 538}

\bibitem{}{Bunker, A.J., Stern, D., Spinrad, H., Dey, A., 
        Steidel, C., \& Wolfe, A. 1998, in preparation}

\bibitem{}{Calzetti, D., Kinney, A.L., \& Storchi-Bergmann, T. 1994, ApJ, 429, 582}

\bibitem{}{Dey, A., Spinrad, H., Stern, D., Graham, J.R., \& Chaffee, F.H.  1998, ApJ, 498, L93}

\bibitem{}{Dickinson, M.E., \etal\ 1998, in preparation}

\bibitem{}{Giavalisco, M., Steidel, C.,  \& Macchetto, D. 1996 ApJ, 470, 189}

\bibitem{}{Hu, E.M., Cowie, L.L., \& McMahon, R.G. 1998, 502, L99}

\bibitem{}{Fern\'andez-Soto, A., Lanzetta, K., \& Yahil, A. 1998, ApJ, submitted}

\bibitem{}{Lanzetta, K., Yahil, A., \& Fern\'andez-Soto, A. 1996, Nature, 381, 759}

\bibitem{}{Lanzetta, K., 1998, in proceedings of  
the Xth Rencontres de Blois, `The Birth of Galaxies', June 28 -
July 4 1998, Blois, France, in preparation}

\bibitem{}{Leitherer, C. \etal\ 1996, PASP, 108, 996}

\bibitem{}{Madau, P., Ferguson, H.C., Dickinson, M.E., Giavalisco,
M., Steidel, C.C., and Fruchter, A. 1996, MNRAS, 283, 1388}

\bibitem{}{Madau, P., Pozzetti, L., and Dickinson, M. 1998, ApJ, 498, 106}

\bibitem{}{Massey, P., Strobel, K., Barnes, J.V., \& 
Anderson, E. 1988, \apj, 328, 315} 

\bibitem{}{Massey, P. \& Gronwall, C. 1990, \apj, 358, 344}

\bibitem{}{Oke, J.B., Cohen, J.G., Carr, M., Cromer, J., Dingizian, A.,
Harris, F.H., Labrecque, S., Lucinio, R., Schall, W., Epps H.,
and Miller, J. 1995, PASP, 107, 375}

\bibitem{}{Marleau, F.R. \& Simard, L. 1998, ApJ, in press} 

\bibitem{}{Soifer, B.T., Neugebauer, G., Franx, M., Matthews, K. \& Illingworth, G. 
1998, ApJ, in press}

\bibitem{}{Spinrad, H., Stern, D., Bunker, A., Lanzetta,
K., Yahil, A., Pascarelle, S. \& Fern\'andez-Soto, A. 1998, 
in preparation}

\bibitem{}{Steidel, C.C., Giavalisco, M., Pettini, M., Dickinson, M.,
and Adelberger, K.  1996, ApJ, 462, L17}

\bibitem{}{Thommes, E., Meisenheimer, K., Fockenbrock, R., Hippelein, 
H., Roser, H.-J., \& Beckwith, S. 1998, MNRAS, 293, 6}

\bibitem{}{Thompson, R.I., Rieke, M., Schneider, G., Hines, D., \& Corbin,
 M. 1998a, ApJ, 492, L95}

\bibitem{}{Thompson, R.I., \etal\ 1998b, in preparation}

\bibitem{}{Williams, R. E., Blacker, B., Dickinson, M., Dixon, W.V.,
Ferguson, H.C., Fruchter, A.S., Giavalisco, M., Gilliland, R.L., Henry, I.,
Katsanis, R., Levay, Z., Lucas, R.A., McElroy, D.B., Petro, L.,
Postman, M. 1996, AJ, 112, 1335}

\end{thebibliography}
\end{document}